\documentclass[sigconf]{acmart}
\acmConference[EASE 2026]{The 30th International Conference on Evaluation and Assessment in Software Engineering}{10–13 June, 2026}{Glasgow, Scotland, United Kingdom}

\usepackage{tabularx}
\usepackage{array}
\usepackage{makecell}

\newcolumntype{Y}{>{\raggedright\arraybackslash}X}

\usepackage{tikz}
\usetikzlibrary{fit}
\usetikzlibrary{arrows.meta,positioning,fit,calc}

\tikzset{
  commit/.style={circle,draw,inner sep=1.4pt,minimum size=4mm},
  mainline/.style={-Latex,thick},
  feature/.style={-Latex,thick,dashed},
  label/.style={font=\small},
  buggy/.style={commit,fill=orange!20,draw=orange!60!black},
  merge/.style={commit,fill=blue!15,draw=blue!50!black},
}

\usepackage{siunitx}  
\usepackage{listings}
\definecolor{dkgreen}{rgb}{0,0.6,0}
\definecolor{gray}{rgb}{0.5,0.5,0.5}
\definecolor{mauve}{rgb}{0.58,0,0.82}

\usepackage{todonotes}
\usepackage{hyperref}

\lstdefinestyle{mystyle}{
    language=C++,       
    commentstyle=\color{gray},
    numbers=left,       
    numberstyle=\footnotesize,
    numbersep=5pt,
    frame=single,       
    tabsize=4           
}

\newcommand{\resultbox}[1]{%
    \par\addvspace{6pt plus 2pt minus 1pt}
    \noindent\begin{minipage}{\columnwidth}
        \centering
        \begin{tikzpicture}
            \node [fill=black!10, draw=none, rounded corners=6pt, 
                   inner sep=6pt, outer sep=0pt,
                   text width=0.95\linewidth] (box) {#1};
        \end{tikzpicture}
    \end{minipage}
    \par\addvspace{6pt plus 2pt minus 1pt}
}

\newcommand{\framework}{\textsc{PhantomRun}}

\ccsdesc[500]{Software and its engineering~Software maintenance tools}
\ccsdesc[300]{Software and its engineering~Empirical software validation}
\ccsdesc[100]{Software and its engineering~Software testing and debugging}

\copyrightyear{2026}
\acmYear{2026}
\setcopyright{cc}
\setcctype{by}
\acmConference[EASE 2026]{The 30th International Conference on Evaluation and Assessment in Software Engineering}{10–13 June, 2026}{Glasgow, Scotland, United Kingdom}
\acmBooktitle{The 30th International Conference on Evaluation and Assessment in Software Engineering (EASE '26), 10–13 June, 2026, Glasgow, Scotland, United Kingdom}
\acmDOI{}
\acmISBN{}

\begin{document}
\title[Where did we fail? --- Reproducing build failures in embedded open source software]{Where did we fail? --- Reproducing build failures in embedded open source software}

\acmConference[EASE 2026]{Evaluation and Assessment in Software Engineering (EASE) 2026}{June 9-12}{Glasgow, United Kingdom}

\makeatletter

\author{Han Fu}
\affiliation{%
  \institution{Ericsson AB}
  \city{Stockholm}
  \country{Sweden}
}
\affiliation{%
  \institution{KTH Royal Institute of Technology}
  \city{Stockholm}
  \country{Sweden}
}
\email{hfu@kth.se}

\author{Andreas Ermedahl}
\affiliation{%
  \institution{Ericsson AB}
  \city{Stockholm}
  \country{Sweden}
}
\affiliation{%
  \institution{KTH Royal Institute of Technology}
  \city{Stockholm}
  \country{Sweden}
}
\email{andreas.ermedahl@ericsson.com}

\author{Sigrid Eldh}
\affiliation{%
  \institution{Ericsson AB}
  \city{Stockholm}
  \country{Sweden}
}
\affiliation{%
  \institution{Mälardalen University}
  \city{Västerås}
  \country{Sweden}
}
\affiliation{%
  \institution{Carleton University}
  \city{Ottawa}
  \country{Canada}
}
\email{sigrid.eldh@ericsson.com}

\author{Kristian Wiklund}
\affiliation{%
  \institution{Ericsson AB}
  \city{Stockholm}
  \country{Sweden}
}
\email{kristian.wiklund@ericsson.com}

\author{Philipp Haller}
\affiliation{%
  \institution{KTH Royal Institute of Technology}
  \city{Stockholm}
  \country{Sweden}
}
\email{phaller@kth.se}

\author{Cyrille Artho}
\affiliation{%
  \institution{KTH Royal Institute of Technology}
  \city{Stockholm}
  \country{Sweden}
}
\email{artho@kth.se}

\renewcommand{\shortauthors}{Fu et al.}

\begin{abstract}
Due to hardware–software co-development in embedded systems, continuous integration (CI) builds frequently fail because of complex cross-compilation, board configurations, and toolchain constraints. Although CI build logs contain valuable diagnostic information, they are short-lived and difficult to reuse due to heterogeneous runners, toolchains, and log formats. To address these challenges, we present \framework{}, a unified abstraction layer and publicly reusable dataset that standardizes the retrieval, storage, and reproduction of CI build logs and metadata. Across~\num{4628} failing CI runs, we reconstructed 91.8\,\% of builds and preserved execution outcomes in 98\,\% of evaluated cases.

\framework{} provides two core capabilities: retrieving the build log of any commit and faithfully re-executing the corresponding build in a controlled environment. By exposing all build artifacts and metadata in a uniform, machine-readable format, \framework{} enables reproducible and longitudinal studies of CI failures. An empirical evaluation shows that reproduced builds closely match their originals, typically differing only in timestamps or minor nondeterministic reordering, demonstrating the feasibility of large-scale historical CI reconstruction.
\end{abstract}

\keywords{continuous integration, software build, compilation error, log parsing} 

\maketitle
\section{Introduction}\label{introduction}
Continuous integration (CI) is a critical mechanism for maintaining build correctness in embedded software development, where hardware–software co-evolution, cross-compilation toolchains, and platform-specific configurations interact in complex and fragile ways~\cite{fu2022prevalence}. CI pipelines orchestrate heterogeneous build environments, toolchains, and configuration layers, serving as the primary automated safeguard against integration regressions.

Compilation failures represent a common and consequential category of CI breakage in embedded systems. They frequently arise from mismatched dependencies, evolving toolchains, configuration drift, and implicit environment assumptions that are difficult to reproduce locally. Such failures are costly~\cite{olsson2012climbing,lwakatare2016towards}: they block integration pipelines, delay feedback cycles, and prevent downstream testing and analysis. Because compilation failures halt the pipeline at its earliest stage, they act as an immediate indicator of environmental and configuration inconsistencies.

In open-source embedded projects, CI build logs are vital for diagnosing compilation failures, yet they are often short-lived due to retention limits and heterogeneous build environments. Failing builds are especially valuable because they expose mismatches between evolving software components and target hardware configurations, and in our prior work, we show that such failures are highly informative for automated analysis and repair~\cite{fu2026phantomrunautorepaircompilation}. Consequently, this work builds on our prior work and focuses specifically on reconstructing failed CI builds to enable a durable, systematic study of compilation-related breakages across both software and hardware dimensions.

Therefore, we present \framework{}, a unified abstraction layer and publicly reusable dataset that standardizes the retrieval, storage, and reproduction of CI build logs and metadata. We mined over~\num{10000} pull and merge requests from four widely used embedded projects—OpenIPC, STM32, RTEMS, and Zephyr\footnote{\href{https://github.com/OpenIPC/firmware}{OpenIPC}, \href{https://github.com/platformio/platform-ststm32}{STM32}, \href{https://gitlab.rtems.org/rtems/rtos/rtems}{RTEMS}, \href{https://github.com/zephyrproject-rtos/zephyr}{Zephyr}}—and reconstructed~\num{4248} failing CI builds containing compilation errors in controlled containerized environments.

Our evaluation shows that 91.8\,\% of failing builds are successfully reconstructed, and reproduced executions preserve termination outcomes and diagnostic structure with high fidelity, typically differing only in benign nondeterminism such as timestamps or minor log reordering. By transforming short-lived CI failures into durable and reproducible artifacts, \framework{} enables systematic cross-project analysis and supports longitudinal, reproducible research on embedded CI behavior.

Accordingly, we investigate the following research questions:

\textbf{RQ1:}~\textit{How embedded software CI builds be reconstructed outside their original CI environments?}\label{RQ1}

\textbf{RQ2:}~\textit{How faithfully do reconstructed CI builds reproduce the compilation failures observed in the original CI executions?}\label{RQ2}

Our main contributions are:

\begin{enumerate}
  \item \textbf{Unified CI Reconstruction Framework.}
  We introduce \framework{}, a reproducible abstraction layer that enables historical CI builds to be retrieved, standardized, and re-executed outside their native CI infrastructures, achieving 91.8\,\% large-scale reconstruction success across heterogeneous embedded CI environments.

  \item \textbf{Large-Scale Reconstruction and Fidelity Evaluation.}
  We reconstruct~\num{4248} failing CI builds across four embedded projects and empirically quantify reconstructability and behavioral fidelity, preserving execution outcomes in 98\,\% of cases and structural log characteristics in 93.6\,\%.
  
  \item \textbf{Publicly Reusable Dataset for Reproducible CI Research.}
  We release a systematically constructed dataset of reproduced CI build logs, metadata, and normalized diagnostics, enabling longitudinal and reproducible studies of compilation failures in embedded systems.
\end{enumerate}

The remainder of the paper is organized as follows. Section~\ref{sec:background_related_work} introduces the context and motivation of our research and reviews related work on open-source software (OSS) CI reconstruction and log parsing. Section~\ref{sec:emberCI} outlines the \framework{} design. Section~\ref{sec:project_mining} describes the methodologies for mining embedded system projects. Section~\ref{sec:methodology} outlines research methodologies, while Section~\ref{sec:study_results} presents our experimental results addressing RQ1 to RQ2. Section~\ref{sec:threats2validity} discusses threats to validity, Section~\ref{sec:disucssion} discusses the results, Section~\ref{sec:conclusion} concludes the main findings, and Section~\ref{sec:futurework} concludes with directions for future work.

\section{Background and Related Work}\label{sec:background_related_work}

Continuous integration (CI) accelerates feedback on integration and configuration problems, but embedded CI is harder to stabilize because builds depend on cross-compilation toolchains, hardware-driven configuration layers, and rapidly evolving dependencies. As a result, compilation failures are common and particularly costly because they block later CI stages. Prior work shows that dependency-related issues account for a substantial share of CI failures across settings~\cite{fu2022prevalence}, motivating approaches that can reproduce historical executions and analyze failures under controlled conditions.

\subsection{CI reconstruction}

GitHub Actions (GHA) and GitLab CI are widely used CI platforms in both OSS and industrial infrastructure.\footnote{\url{https://github.com/features/actions}, \url{https://docs.gitlab.com/ci/}} Workflows are defined in YAML and specify runner environments, setup commands, and build steps. GHA workflows live under \texttt{.github/workflows/} and integrate tightly with GitHub’s event model~\cite{github_actions_about_2024}, whereas GitLab CI uses \texttt{.gitlab-ci.yml} executed by GitLab runners with flexible deployment models~\cite{gitlab_ci_docs_2024}. These differences affect what environment details are explicit and therefore what can be reconstructed reliably.

Recent research has explored replaying CI executions for reproducible studies, including workflow reconstruction and replay~\cite{DBLP:conf/wcre/GolzadehDM22,DBLP:conf/icse/ZhuGFR23}, mining workflow histories for recurring issues~\cite{DBLP:conf/msr/ValenzuelaToledoBKN23}, and replaying bug-fix scenarios to study CI reliability~\cite{DBLP:conf/icse/SaavedraSM24}. However, cross-platform reconstruction that targets embedded compilation behavior and supports structured, log-based comparisons remains limited, motivating our focus on unified reconstruction across GHA and GitLab CI.

\subsection{Hardware-in-the-loop with dependency errors}

Embedded CI often interacts with hardware constraints (e.g., proprietary SDKs, device availability, or hardware-in-the-loop setups), which can make CI adoption and automation fragile~\cite{olsson2012climbing,lwakatare2016towards}. Dependency inconsistencies further amplify this fragility: mismatched toolchains, BSPs, middleware versions, and configuration assumptions frequently manifest as compilation failures~\cite{kerzazi2014automated,fu2022prevalence}. Related work highlights broader consequences such as portability limits, configuration drift, and dependency decay~\cite{fischer2020forgotten,khazem2018making,zakaria2022mapping,ratti2018conceptual,DBLP:conf/sigsoft/0001RJGA19}. These findings collectively motivate studying whether historical embedded CI failures can be reconstructed and compared faithfully despite environment drift.

\subsection{Empirical log parsing for CI analysis}

CI logs capture compiler diagnostics, linker output, and build-system traces, but they are large, heterogeneous, and costly to analyze manually~\cite{rodrigues2021clp}. Automated log analysis techniques (e.g., LogRAM, DeepLog, and LogTools) demonstrate how unstructured logs can be transformed into structured representations at scale~\cite{dai2020logram,du2017deeplog,zhu2019tools}. Log mining has also been used for execution tracing and fault localization~\cite{fu2014digging,he2018identifying,zhang2021onion}, suggesting that structured log signals can support systematic failure analysis beyond ad hoc inspection.

Empirical evaluations of log parsers on public datasets further document how parser choices and log characteristics affect extraction quality~\cite{DBLP:conf/dsn/HeZHLL16,DBLP:conf/icse/ZhuHLHXZL19}. CI build logs differ from many commonly studied system logs in that they represent short-lived, commit-specific build executions that aggregate outputs from compilers, build systems, dependency managers, and test runners into a single stream. We therefore apply automated parsing and normalization to extract comparable diagnostic structure and enable quantitative evaluation of reconstruction fidelity across platforms.

\subsection{Log-Based CI Failure Analysis}

Automated log parsing has been extensively studied as a means to transform unstructured textual logs into structured event templates suitable for analysis. Existing approaches span rule-driven methods~\cite{DBLP:journals/apin/DamasioFNPS02}, static source-based techniques that extract logging patterns directly from code~\cite{DBLP:conf/icml/XuHFPJ10,DBLP:conf/issre/NagappanWV09}, and a broad family of data-driven methods based on clustering, heuristic grouping, or frequent-pattern mining~\cite{DBLP:conf/qsic/JiangHFH08,DBLP:journals/tkde/MakanjuZM12,DBLP:conf/icws/HeZZL17,DBLP:conf/icdm/FuLWL09,DBLP:conf/cikm/TangLP11,DBLP:conf/cnsm/VaarandiP15,vaarandi2003data,DBLP:journals/tse/DaiLCSC22}. Learning-based approaches further extend this line of work by leveraging sequence modeling or neural representations to infer log templates automatically~\cite{du2017deeplog,zhu2019tools,DBLP:conf/www/LiuZHZLKXMLDRZ22}. 

While these techniques demonstrate the feasibility of large-scale log structuring, they have primarily been evaluated on system logs or distributed service traces rather than CI build outputs. CI logs differ in several respects: they contain heterogeneous tool invocations, nested build phases, compiler diagnostics, and environment-dependent artifacts that may vary across platforms and runners. Moreover, CI logs are tightly coupled with versioned source code and configuration files, introducing additional context not typically considered in generic log parsing studies.

Log mining has also been applied to execution tracing and fault localization~\cite{fu2014digging,he2018identifying,zhang2021onion}, showing that structured log representations can support automated reasoning about failure causes. However, systematic studies focusing on compilation failures in CI pipelines—particularly in embedded software contexts—remain limited. In this work, we build upon automated parsing principles to normalize CI build logs and extract comparable diagnostic structures. These structured representations enable quantitative comparison between original and reconstructed executions, forming the basis for evaluating reconstruction fidelity across heterogeneous CI platforms.

\section{\framework{} Setup}\label{sec:emberCI}
\begin{figure}[htbp]
    \centering  
    \includegraphics[scale=.95]{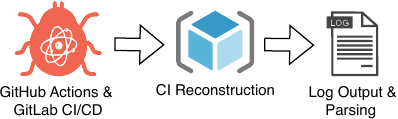}
    \Description{A diagram illustrating the whole process of docker images rebuild}
    \caption{CI rebuild process}
    \label{fig:Docker_images_rebuild}
\end{figure}

We present \framework{}, a fully automated CI rebuild framework for studying compilation failures in open-source embedded software. As shown in Fig.~\ref{fig:Docker_images_rebuild}, the framework reconstructs the build environment of failed GitHub pull requests and GitLab merge requests inside isolated Docker containers, and re-executes the corresponding build steps without modifying the project’s CI configuration. Each reconstructed build is then rerun to regenerate the build log, from which compilation errors and related diagnostics are parsed and normalized into a structured representation. This non-intrusive, project-agnostic pipeline supports heterogeneous C/C++ embedded projects and provides a reusable basis for reproducible CI failure analysis, dataset construction, and downstream empirical evaluation.

\subsection{CI Reconstruction Framework}\label{subsec:ci_reconstruction_framework}
The CI reconstruction framework is designed as a modular pipeline that supports the discovery, reconstruction, and analysis of compilation errors from CI executions. Its components can be executed independently or composed end-to-end, enabling flexible reuse across projects, CI platforms, and analysis tasks. In this study, the framework is used to reconstruct CI builds, recover their associated logs, and produce stable artifacts suitable for empirical analysis.

For each failed CI run, the framework first establishes a stable source-code baseline by retrieving the corresponding repository state and checking out the commit associated with the failure. This anchoring step ensures that subsequent reconstruction and analysis operate on the exact source version evaluated by the original CI system. The framework explicitly accounts for different pull request and merge request integration strategies, which influence how failing and succeeding code states are represented in version control.

Given the selected repository state, the framework reconstructs the CI build environment in an isolated, containerized setting. CI configuration files and referenced scripts are analyzed to identify the operating system, toolchain setup, environment variables, and build commands relevant to compilation. These elements are used to re-execute the compilation stage of the CI workflow under controlled conditions, producing raw build logs without modification or interpretation.

Build logs are obtained either from reconstructed executions or directly from CI artifacts provided by the hosting platform. Treating both sources uniformly allows the framework to analyze historical failures even when full reconstruction is unnecessary or partially infeasible. All logs are stored as immutable artifacts and passed to a dedicated parsing stage that extracts compilation-related diagnostics and normalizes them into a structured representation suitable for cross-project analysis.

\begin{figure}[htbp]
    \centering  
    \includegraphics[scale=.9]{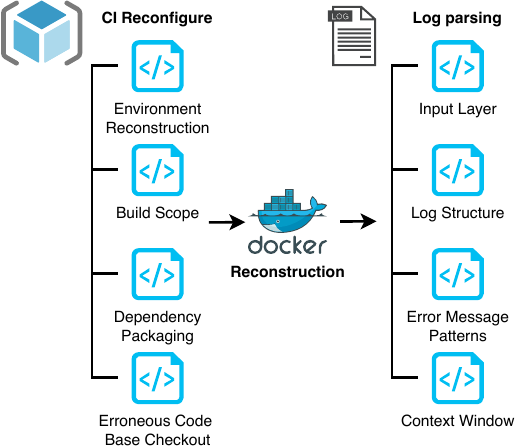}
    \Description{A diagram showing the process of how to reproduce a CI failed run.}
    \caption{CI reconstruction and log parsing}
    \label{fig:CI_Docker}
\end{figure}

\subsection{Properties of Reconstructed CI Logs}\label{subsec:ci_reconstruction_properties}

To support reproducible and comparative analysis, CI reconstruction is guided by a set of properties that characterize the validity of reconstructed build logs as empirical data.

\textbf{Environment determinism.}
Reconstructed builds preserve the environment relevant to compilation diagnostics, including toolchains, configuration parameters, and build commands specified in CI workflows. Executing builds in isolated environments reduces variability introduced by external system state and ensures that compilation behavior is driven primarily by the analyzed code and configuration.

\textbf{Stage isolation.}
Reconstruction is intentionally restricted to the compilation stage of CI pipelines. This isolates the source of build failures from downstream stages such as testing or deployment, which depend on built artifacts and may introduce additional sources of nondeterminism.

\textbf{Diagnostic preservation.}
Rather than reproducing logs verbatim, the framework prioritizes preserving the semantic content of compilation diagnostics. Compiler and build-system error messages, file references, and contextual information are retained, while superficial differences such as formatting or auxiliary output are abstracted away.

\textbf{Cross-project comparability.}
Reconstructed logs originate from projects with diverse CI infrastructures and build systems. By normalizing diagnostics into a structured representation, the framework abstracts away project-specific logging conventions while retaining compiler-relevant information, enabling consistent analysis across heterogeneous projects.

\textbf{Stability under nondeterminism.}
We define stability as the property that repeated reconstructions of the same CI build produce identical compilation outcomes and semantically equivalent diagnostics. While superficial variations (e.g., timestamps, runner-specific paths, or parallel step ordering) may differ across executions, the presence, type, and location of compilation errors remain invariant. This ensures that reconstructed logs serve as reliable empirical artifacts despite benign execution-level nondeterminism.

In the raw datasets, contextual dimensions are implicitly encoded in log filenames using project-specific naming schemes. Representative forms include:

\begin{verbatim}
pr-<id>.log
pr-<id>__<target>.log
proj<project-id>_mr<mr-id>_sha<commit>.log
*_Firmware(<hardware>).log
\end{verbatim}

CI logs frequently contain runner-specific absolute paths that require prefix normalization before comparison. Log filenames further encode project-specific metadata—such as integration identifiers, commit SHAs, hardware variants, or matrix-expanded targets—using non-uniform naming schemes that must be standardized for cross-project analysis. A more substantial challenge is the heterogeneity of build inputs: different container environments, build systems, and compilers generate distinct output formats and diagnostics. Our framework addresses this diversity through tool-aware parsing and normalization to produce comparable representations across CI environments.

\section{Project Mining}\label{sec:project_mining}
Our pipeline automates the discovery of embedded repositories and the extraction of compilation-failure logs from their CI histories. The workflow consists of four stages:

\begin{itemize}
    \item \textbf{Authentication:} Authenticate to the GitHub and GitLab APIs using a personal access token (PAT) to access metadata at higher rate limits.
    \item \textbf{Repository discovery:} Query for repositories tagged with \texttt{embedded}, primarily written in C or C++, and exceeding a popularity threshold (Stars $>20$).
    \item \textbf{CI inspection:} Retrieve CI workflow runs for each repository and filter to runs with a \texttt{failure} conclusion.
    \item \textbf{Log extraction:} Download log streams and apply regular expressions to detect compilation-specific diagnostics (e.g., \texttt{fatal error}, \texttt{undefined reference}).
\end{itemize}

\subsection{Project Selection Strategy}
To construct a representative dataset of embedded compilation failures, we targeted repositories that jointly satisfy domain relevance, language suitability, and a minimal maturity threshold:

\begin{itemize}
    \item \textbf{Domain relevance:} The repository is tagged with \texttt{embedded}.
    \item \textbf{Language constraint:} The primary language is C/C++, reflecting common practice in embedded firmware development.
    \item \textbf{Project maturity:} More than 20 stars to reduce toy projects while retaining active, community-used codebases.
\end{itemize}

The resulting query yields a candidate set of repositories $R$, which we then analyze via CI metadata and logs.

\subsection{Data Collection and Artifact Mining}
For each candidate repository $r \in R$, we mine CI metadata and retain only runs that correspond to build workflows and end in failure. Formally, we keep runs satisfying:

\begin{equation}
    R_{valid} = \{\, r \in R \mid \text{fail}(r) \land \text{build\_workflow}(r) \,\}
\end{equation}

This filtering excludes failures unrelated to compilation (e.g., documentation builds, formatting checks, or bot automation). The pipeline paginates through API results, extracts the repository identifier (Owner/Name), and queries endpoints such as \texttt{/actions/runs} to enumerate workflow history and retrieve log artifacts.

Applying these filters produced approximately 60 candidate repositories. From these, we selected Zephyr, OpenIPC, and STM32 for detailed analysis. In parallel, we performed automated mining on public GitLab instances, including the RTEMS GitLab server. For each project, data collection spans from the first available pull/merge request up to October~3,~2025; later requests are outside our study window.

\subsection{Project specifications}\label{subsubsec:project_specifications}

Tables~\ref{tab:projects_sys_arch} and~\ref{tab:projects_ci_build_failures} summarize the four studied projects and motivate their selection as a diverse set of embedded CI systems. RTEMS and Zephyr represent mature, widely deployed RTOSes but differ sharply in tooling and CI organization. As shown in Table~\ref{tab:projects_sys_arch}, RTEMS emphasizes long-term stability and a traditional RTOS stack, while Zephyr reflects a large, modular system with extensive configuration layers (CMake, Kconfig, devicetree, and \texttt{west}). These differences are directly reflected in CI practice (Table~\ref{tab:projects_ci_build_failures}): RTEMS builds are centered on autotools and \texttt{make}, producing largely linear logs that progress from configuration to compilation and linking (configure, make, GCC/LD). In contrast, Zephyr relies on a multi-stage toolchain consisting of \texttt{west}, CMake, Ninja, and the compiler/linker, and generates structured logs through the \texttt{twister} test harness with automated QEMU execution. Over the same six-month window, we collected comparable volumes of compilation failures (\num{1784} for RTEMS and \num{1989} for Zephyr), but with different dominant failure sources (e.g., configure/Makefile/linker scripts versus CMake/Kconfig/devicetree and conditional compilation diagnostics).

\begin{table*}
\renewcommand{\arraystretch}{1.2}
\caption{Project Comparison: System and Architecture}
{\small
\begin{tabularx}{\textwidth}{@{}lYYYY@{}}
\toprule
\textbf{Aspect} & \textbf{RTEMS} & \textbf{Zephyr} & \textbf{OpenIPC} & \textbf{STM32} \tabularnewline
\midrule

\multicolumn{5}{c}{\textbf{Overview}} \tabularnewline

\textbf{Platforms}
& GitLab
& GitHub
& GitHub
& GitHub \tabularnewline

\textbf{License}
& BSD-2-Clause (GPL)
& Apache 2.0
& GPLv3
& Apache 2.0 \tabularnewline

\textbf{Governance}
& OAR Corp, OSU-hosted
& Linux, TSC-based
& Community-driven
& PlatformIO organization \tabularnewline

\textbf{Community}
& Centralized; expert-driven
& Large-scale; Linux Foundation TSC governance
& Community-driven; enthusiast contributors
& Broad developer base \tabularnewline

\midrule

\multicolumn{5}{c}{\textbf{Architecture}} \tabularnewline

\makecell[l]{\textbf{Supported}\\\textbf{Architectures}}
& ARM, MIPS, PowerPC, SPARC, RISC-V, x86
& ARM, x86, RISC-V, ARC, Xtensa, Nios II
& ARM Cortex-A (HiSilicon, XM, Fullhan SoCs)
& STM32 (ARM Cortex-M0/M3/M4/M7) \tabularnewline

\textbf{Kernel Type}
& Monolithic RTOS
& Modular RTOS
& Linux (monolithic)
& N/A \tabularnewline

\makecell[l]{\textbf{Memory}\\\textbf{Management}}
& No MMU; static allocation
& MPU/MMU support; user-space isolation
& Linux MMU-based (platform-dependent)
& N/A (build platform) \tabularnewline

\makecell[l]{\textbf{Multicore}\\\textbf{Support}}
& SMP support; deterministic synchronization (MrsP)
& SMP support; IPC mechanisms
& Platform-dependent (Linux SMP support)
& N/A \tabularnewline

\makecell[l]{\textbf{Real-Time}\\\textbf{Scheduling}}
& Fixed-priority preemptive scheduling
& Configurable preemptive/cooperative scheduling
& Provided by underlying Linux kernel
& Depends on selected RTOS (e.g., FreeRTOS) \tabularnewline

\makecell[l]{\textbf{Power}\\\textbf{Management}}
& Basic idle and low-power modes
& Dynamic power management; tickless kernel; per-device PM
& Provided by Linux kernel and SoC
& Depends on STM32 HAL and middleware \tabularnewline

\textbf{Build System}
& Waf + RTEMS Source Builder (RSB)
& West + CMake + Kconfig + Devicetree + Ninja
& OpenWrt Makefiles + Buildroot-based tooling
& PlatformIO Core (Python/SCons) \tabularnewline

\textbf{OS / Language}
& RTOS / C
& RTOS / C, C++
& Linux-based firmware / C
& Build platform / Python, C++ \tabularnewline

\textbf{APIs \& Standards}
& RTEMS Classic API; POSIX; FreeBSD networking stack
& POSIX APIs; RTOS primitives; Devicetree overlays
& OpenWrt userland; BusyBox
& Arduino; CMSIS; STM32Cube HAL; SPL \tabularnewline

\textbf{Networking}
& FreeBSD TCP/IP; NFS; FAT
& IPv4/6; BLE; 802.15.4; MQTT; CoAP; CAN; USB
& Full TCP/IP; ONVIF; RTSP; Wi-Fi
& Application-dependent (via STM32 libraries) \tabularnewline
\midrule

\multicolumn{5}{c}{\textbf{Performance and Real-Time Characteristics}} \tabularnewline

\textbf{Performance}
& Deterministic real-time execution
& \textasciitilde500-cycle context switch; efficient IPC
& SoC-dependent; optimized for video streaming workloads
& MCU- and RTOS-dependent performance \tabularnewline

\makecell[l]{\textbf{Tooling \&}\\\textbf{CI}}
& Custom toolchain
& West CLI
& OpenWrt build system
& PlatformIO CLI; VSCode \tabularnewline
\tabularnewline
\midrule

\multicolumn{5}{c}{\textbf{Infrastructure}} \tabularnewline

\makecell[l]{\textbf{Vendor}\\\textbf{Infrastructure}}
& Aerospace-focused vendors
& Broad industry (Intel, Nordic, NXP, ST, etc.)
& SoC vendors (HiSilicon, XM, Fullhan)
& Official STMicroelectronics \tabularnewline

\makecell[l]{\textbf{Release}\\\textbf{Cadence}}
& Conservative; long-term stability focused
& Active; regular (quarterly+) releases
& Irregular; community-driven updates
& Regular; aligned with PlatformIO releases \tabularnewline
\bottomrule
\end{tabularx}}
\label{tab:projects_sys_arch}
\end{table*}

\begin{table*}
\renewcommand{\arraystretch}{1.2}
\caption{Project Comparison: CI Tooling, Build, and Failure Characteristics}
{\small
\begin{tabularx}{\textwidth}{@{}lYYYY@{}}
\toprule
\textbf{Aspect} & \textbf{RTEMS} & \textbf{Zephyr} & \textbf{OpenIPC} & \textbf{STM32} \tabularnewline
\midrule

\multicolumn{5}{c}{\textbf{CI Tooling and Developer Workflow}} \tabularnewline

\makecell[l]{\textbf{CI}\\\textbf{Infrastructure}}
& BuildBot; shell-based workflows
& CMake-driven CI matrix
& Community-maintained scripts
& PlatformIO CLI-based workflows \tabularnewline

\textbf{CI Platforms}
& GitLab CI (MR)
& GitHub Actions (PR)
& GitHub Actions (PR)
& GitHub Actions (PR) \tabularnewline

\textbf{Test Framework}
& Custom runners; QEMU BSPs
& \texttt{twister} test harness
& Custom scripts; hardware-based validation
& PlatformIO integrated testing framework \tabularnewline

\textbf{QEMU Integration}
& Manual configuration
& Fully automated via \texttt{twister}
& N/A
& Limited; hardware-based testing predominant \tabularnewline

\makecell[l]{\textbf{CI Board}\\\textbf{Coverage}}
& $\sim$10--15 boards
& 50+ QEMU boards; selected physical hardware
& Selected SoCs with predefined configurations
& STM32 boards (official and community-supported) \tabularnewline

\textbf{Packaging \& SBOM}
& Manual toolchain management; no formal packaging
& Optional SBOM generation; OTA artifacts supported
& N/A
& Partial SBOM support via PlatformIO \tabularnewline

\makecell[l]{\textbf{Static}\\\textbf{Analysis}}
& Minimal; manual execution
& \texttt{clang-format}, \texttt{cppcheck}, MISRA, SPDX via CI
& Manual or project-specific tooling
& Integrates \texttt{cppcheck}, \texttt{clang-tidy} via PlatformIO \tabularnewline

\makecell[l]{\textbf{Code}\\\textbf{Coverage}}
& Manual scripts with GCC instrumentation
& Integrated GCOV via \texttt{twister}
& N/A
& GCOV support via PlatformIO test runners \tabularnewline

\makecell[l]{\textbf{CI Matrix}\\\textbf{Flexibility}}
& Basic shell-based matrix; single-host execution
& Multi-axis YAML matrix (arch $\times$ board $\times$ features)
& Manual configuration scripting
& Flexible YAML matrix for boards and environments \tabularnewline

\makecell[l]{\textbf{Compilation Errors}\\\textbf{(6 months)}}
& 1784
& 1989
& 31
& 444 \tabularnewline[2pt]

\textbf{Build Tools}
& \texttt{make} with autotools
& \texttt{west} + CMake + Ninja
& \texttt{make} with Buildroot
& Python-based SCons \tabularnewline

\textbf{Infrastructure}
& Self-hosted GitLab instance; custom build system
& Modular kernel; CMake + Kconfig; \texttt{west}-managed
& Vendor SDKs; custom cross-toolchains
& JSON-based board configs; PlatformIO build scripts \tabularnewline
\midrule

\multicolumn{5}{c}{\textbf{Failure Characteristics}} \tabularnewline

\textbf{Log Structure}
& Autotools-driven (configure $\rightarrow$ make $\rightarrow$ GCC/LD)
& CMake/Ninja pipeline (CMake $\rightarrow$ Ninja $\rightarrow$ GCC/LD)
& Make/Buildroot pipeline (make $\rightarrow$ Buildroot $\rightarrow$ cross-GCC/LD)
& PlatformIO/SCons pipeline (SCons $\rightarrow$ GCC/Clang $\rightarrow$ LD) \tabularnewline

\makecell[l]{\textbf{Typical Failure}\\\textbf{Sources}}
& Autoconf macros; Makefile rules; linker scripts
& CMake, Kconfig, devicetree; compiler and linker diagnostics
& Toolchain configuration; packaging; kernel configuration
& Python execution; SCons tasks; compiler/linker errors \tabularnewline

\makecell[l]{\textbf{Error Pattern}\\\textbf{Coverage}}
& Compiler, linker, configure, make
& Compiler, linker, CMake, Ninja, Kconfig, \texttt{west}
& Make, Buildroot, cross-compiler, packaging
& Compiler, linker, SCons, Python exceptions \tabularnewline

\textbf{Example Error}
& \texttt{arm-rtems6-gcc: undefined reference to 'rtems\_task\_create'}
& \texttt{src/main.c:42:17: error: 'foo' undeclared}
& \texttt{make[2]: *** [package] Error 2}
& \texttt{src/main.cpp:12:10: fatal error: Arduino.h: No such file or directory} \tabularnewline

\bottomrule
\end{tabularx}}
\label{tab:projects_ci_build_failures}
\end{table*}

To broaden beyond RTOS-centric development, we also include OpenIPC and STM32. Table~\ref{tab:projects_sys_arch} highlights that OpenIPC is firmware-oriented (Linux/OpenWrt-style userland on ARM Cortex-A), while STM32 is tooling-centric (PlatformIO Core and SCons supporting many STM32 targets). Their CI and failure characteristics (Table~\ref{tab:projects_ci_build_failures}) complement the RTOS projects: OpenIPC logs are aggregated around Make/Buildroot and failures often stem from toolchain and packaging/configuration issues, whereas STM32 logs are layered around PlatformIO/SCons and include Python/SCons task failures alongside compiler/linker errors. Although their failure volumes differ substantially (31 for OpenIPC and 444 for STM32), together the four projects span distinct CI platforms, build stacks, and log structures, enabling a more general evaluation of reconstruction feasibility and behavioral fidelity across heterogeneous embedded CI environments.

\section{Methodology}\label{sec:methodology}
To answer the two research questions, we structure our methodology around \framework{} regarding the lifecycle of a CI failure: reconstructing the CI build and diagnosing the compilation error, and assessing the fidelity of the reproduced execution. All experiments are conducted on the systematically constructed dataset described in Section~\ref{sec:project_mining}, spanning four embedded software projects and multiple CI platforms.

\subsection{Methodology for RQ1: CI Build Reconstruction}\label{subsec:method_rq1}

RQ1 investigates whether embedded software CI builds can be reconstructed outside their original CI environments. The reconstruction process reuses the original CI specifications to extract the runner operating system, environment variables, setup commands, and build steps, which are translated into a Dockerfile—a declarative specification for building a containerized execution environment—and an executable build script. Each reconstructed build is executed inside this isolated container environment, independently of the original CI infrastructure and without manual intervention.

We consider a CI build successfully reconstructed if the containerized execution completes the intended build stage and produces a build outcome (success or failure). The reconstruction success rate across projects and CI platforms serves as the primary metric for RQ1.

\subsection{Methodology for RQ2: Reconstruction Fidelity}\label{subsec:method_rq2}

RQ2 evaluates the \emph{reconstruction fidelity} of containerized CI replay. We define reconstruction fidelity as the extent to which a reconstructed build preserves the observable compilation behavior of the original CI execution. Observable behavior is evaluated along two hierarchical dimensions:

\begin{enumerate}
    \item \textbf{Outcome equivalence (behavioral fidelity).} 
    Each run is classified by its final compilation status (Success or Failure). Outcome equivalence holds when the reconstructed execution and the original CI run produce the same compilation outcome.

    \item \textbf{Diagnostic structure equivalence (structural fidelity).} 
    The original CI log and the reconstructed log are parsed and normalized using the pipeline described in Section~\ref{subsec:ci_reconstruction_framework}. This produces a structured representation capturing compiler diagnostic categories, referenced source files and locations, and build-stage boundaries. Diagnostic structure equivalence holds when these normalized representations are identical.
\end{enumerate}

Outcome equivalence captures coarse-grained behavioral consistency (i.e., whether the build succeeds or fails), whereas diagnostic structure equivalence captures fine-grained structural consistency of compiler diagnostics and build-stage organization. Diagnostic structure equivalence is therefore a strictly stronger criterion than outcome equivalence.

From the~\num{4248} successfully reconstructed builds identified in RQ1, we draw a stratified proportional sample of 50 builds (1 from OpenIPC, 5 from STM32, 21 from RTEMS, and 23 from Zephyr). Each sampled build is paired with its corresponding original CI log and evaluated independently along the two fidelity dimensions.

Scoring is binary: for each build and each criterion, equivalence either holds or does not. No partial credit is assigned. Reconstruction fidelity is reported as the proportion of sampled builds satisfying each equivalence criterion.

\section{Result}\label{sec:study_results}

\subsection{RQ1: Reconstructability of Embedded CI Builds}\label{subsec:rq1_results}

To answer RQ1, we attempted reconstruction for all~\num{4628} failing CI runs across the four projects.

As shown in Table~\ref{tab:rq1_reconstruction},~\num{4248} builds were successfully reconstructed, yielding an overall success rate of 91.8\,\%. Reconstruction rates exceeded 89\,\% for every project, reaching 96.8\,\% for OpenIPC, 94.1\,\% for STM32, 92.4\,\% for Zephyr, and 89.7\,\% for RTEMS. Achieving over 90\,\% replay success is notable given the heterogeneity of CI infrastructures, container images, toolchains, and external dependencies involved in embedded builds. In total, 380 builds could not be reconstructed: 1 in OpenIPC, 11 in STM32, 164 in Zephyr, and 204 in RTEMS. These results indicate that embedded CI failures can be deterministically replayed outside their native CI environments at scale despite substantial environmental variability.

\begin{table}[t]
\caption{Reconstruction Success per Project}
\centering
\small
\renewcommand{\arraystretch}{1.1}
\setlength{\tabcolsep}{6pt}
\begin{tabular}{l r r@{\hspace{4pt}}r}
\toprule
\textbf{Project} & \textbf{Attempts} & \multicolumn{2}{c}{\textbf{Reconstructed (\%)}} \\
\midrule
OpenIPC & 32   & 31   & (96.8\,\%) \\
STM32   & 455  & 444  & (94.1\,\%) \\
RTEMS   & 1988 & 1784 & (89.7\,\%) \\
Zephyr  & 2153 & 1989 & (92.4\,\%) \\
\midrule
\textbf{Total} & 4628 & 4248 & (91.8\,\%) \\
\bottomrule
\end{tabular}
\label{tab:rq1_reconstruction}
\end{table}

Table~\ref{tab:rq1_failure_causes} breaks down the 380 reconstruction failures. Missing hardware dependencies account for 34\,\% (129 cases), removed or outdated package repositories for 27\,\% (103 cases), unavailable proprietary toolchains for 19\,\% (72 cases), and implicit environment dependencies not captured in CI scripts for 15\,\% (57 cases). The remaining 5\,\% (19 cases) fall into miscellaneous causes. The distribution shows that reconstruction failures predominantly stem from external dependency decay and underspecified build environments rather than from inherent limitations of the container-based reconstruction approach.

\begin{table}[t]
\caption{Primary Causes of Reconstruction Failure}
\centering
\small
\begin{tabular}{l r r@{\hspace{4pt}}}
\toprule
\textbf{Failure Cause} & \textbf{Count} & \textbf{Percentage} \\
\midrule
Hardware dependency missing & 129 & 34\,\% \\
Removed package repository & 103 & 27\,\% \\
Proprietary toolchain unavailable & 72 & 19\,\% \\
Implicit environment dependency & 57 & 15\,\% \\
Other & 19 & 5\,\% \\
\midrule
\textbf{Total} & 380 & 100\,\% \\
\bottomrule
\end{tabular}
\label{tab:rq1_failure_causes}
\end{table}

Overall, the results demonstrate that large-scale preservation and replay of embedded CI compilation failures is technically feasible, with most unreconstructed cases attributable to external dependency volatility rather than methodological constraints.

\resultbox{
    \textbf{Finding 1:} Reconstruction of embedded CI compilation failures is broadly feasible: most failing builds can be deterministically replayed, with remaining cases primarily limited by external dependency volatility rather than technical constraints.}\label{result:finding1}

\subsection{RQ2: Reconstruction Fidelity}\label{subsec:rq2_results}

We evaluated reconstruction fidelity on the stratified sample of 50 builds (Section~\ref{subsec:method_rq2}). 

\textbf{Outcome equivalence} was satisfied in 49 of 50 builds (98\,\%), indicating that containerized replay almost always preserves the final compilation status of the original CI execution.

\textbf{Diagnostic structure equivalence} was satisfied in 47 of 50 builds (94\,\%), indicating that in most cases, reconstructed builds also preserve the normalized compiler diagnostic structure and build-stage organization.

Table~\ref{tab:rq2_fidelity} reports per-project equivalence rates, computed as the fraction of sampled builds within each project satisfying each criterion. Percentages are derived directly from the project-specific sample sizes.

Cases failing diagnostic structure equivalence generally preserved outcome equivalence and differed only in non-critical warnings, auxiliary build steps, or minor environment drift (e.g., implicit compiler version changes or nondeterministic ordering not eliminated by normalization). Complete behavioral mismatches were rare and attributable to subtle dependency or toolchain differences.

\begin{table}[t]
\caption{Reconstruction Fidelity per Project}
\centering
\small
\begin{tabular}{l r r}
\toprule
\textbf{Project} & \textbf{Outcome} & \textbf{Structure} \\
\midrule
RTEMS (n=21) & 20/21 (95.2\,\%) & 19/21 (90.5\,\%) \\
Zephyr (n=23) & 23/23 (100\,\%) & 22/23 (95.7\,\%) \\
OpenIPC (n=1) & 1/1 (100\,\%) & 1/1 (100\,\%) \\
STM32 (n=5) & 5/5 (100\,\%) & 5/5 (100\,\%) \\
\midrule
\textbf{Overall (n=50)} & 49/50 (98\,\%) & 47/50 (94\,\%) \\
\bottomrule
\end{tabular}
\label{tab:rq2_fidelity}
\end{table}

\resultbox{
    \textbf{Finding 2:} Reconstructed CI builds preserve both execution outcomes and diagnostic structure with high fidelity, indicating that containerized replay reliably captures the observable behavior of embedded CI executions.
}\label{result:finding2}

\section{Threats to Validity}\label{sec:threats2validity}
\noindent\textbf{Internal Validity.}
Discrepancies may arise if CI configurations or build steps are incompletely captured during reconstruction. We mitigate this risk by systematically extracting build specifications and executing all reconstructed builds in isolated, containerized environments.

\noindent\textbf{External Validity.}
Our findings are based on four open-source embedded projects using GitHub Actions and GitLab CI. While these projects cover diverse build systems and CI practices, the results may not generalize to proprietary CI infrastructures, non-embedded domains, or projects relying on substantially different tooling.

\noindent\textbf{Construct Validity.}
We measure reconstruction success and fidelity using build outcomes and compilation diagnostics extracted from CI logs. These measures capture practical reproducibility but may abstract away finer-grained differences, such as performance variation or nondeterministic log artifacts.

\noindent\textbf{Conclusion Validity.}
Our conclusions are based on observed reconstruction success rates and log fidelity across the collected CI runs. While these results indicate consistent reconstruction behavior, they are limited to the studied projects and time period. Larger datasets and additional projects would further strengthen the conclusions.

\section{Discussion}\label{sec:disucssion}

Across~\num{4628} failing CI runs, we reconstructed 91.8\,\% of builds and preserved execution outcomes in 98\,\% of evaluated cases. Together, RQ1 and RQ2 demonstrate that embedded CI failures are reconstructable at scale and behaviorally faithful under containerized replay. Although CI logs are typically transient artifacts of diverse and evolving CI environments, our results show that they can be preserved as durable and reproducible research assets.

Reconstruction failures were mainly caused by changes in external dependencies rather than methodological limitations. Missing hardware components, removed repositories, proprietary toolchains, and minor environment drift accounted for most unreconstructed cases. These results show that long-term CI reproducibility is primarily constrained by external dependency changes.

\section{Conclusion}\label{sec:conclusion}

Build logs from continuous integration (CI) systems contain valuable information about compilation behavior and failure causes, yet they are typically short-lived and difficult to reuse due to heterogeneous runners, toolchains, and infrastructure dependencies. To address these challenges, base on our prior work~\cite{fu2026phantomrunautorepaircompilation}, we present \framework{}, a unified abstraction layer and publicly reusable dataset that standardizes the retrieval, storage, and reproduction of CI build logs and metadata.

Through large-scale evaluation across four embedded systems projects, we show that CI failures are reconstructable at scale and behaviorally faithful under containerized replay. We successfully reconstructed~\num{4248} failing CI runs of 91.8\,\% succesful rate, and reproduced execution outcomes in 98\,\% of evaluated cases, with structural fidelity preserved in the vast majority of builds. Reconstruction failures were primarily attributable to external dependency volatility and external dependency decay rather than methodological constraints.

These findings demonstrate that embedded CI builds can transition from ephemeral operational artifacts to durable, reproducible research assets. By normalizing heterogeneous CI environments into a consistent abstraction, \framework{} enables reliable historical reconstruction and supports reproducible, longitudinal studies of compilation failures in embedded systems.

\section{Future Work}\label{sec:futurework}
This work opens several directions for extending CI reconstruction and analysis beyond compilation failures. First, we plan to extend CI reconstruction beyond the compilation phase to include testing and integration stages. Second, we intend to apply advanced AI techniques to support a higher-level understanding of reconstructed CI logs. Rather than focusing on failure resolution, such techniques can be used to summarize complex build logs, correlate diagnostics across stages, and identify recurring failure patterns at scale. Finally, we plan to explore the applicability of the proposed CI reconstruction approach in industrial settings, extending the framework to proprietary codebases and enterprise CI infrastructures to assess its robustness and support for industrial debugging and reliability analysis.

\section*{Data Availability}
To facilitate the replication of our results and support further research, we have made our dataset and generation scripts available on Zenodo at: \href{https://doi.org/10.5281/zenodo.19894284}{\underline{\textcolor{blue}{EASE-Repo}}}.

\begin{acks} 
This work was partially supported by the Wallenberg Artificial Intelligence, Autonomous Systems and Software Program (WASP) funded by the Knut and Alice Wallenberg Foundation.
\end{acks}

\bibliographystyle{ACM-Reference-Format}
\balance
\bibliography{references}

\end{document}